\documentclass[twocolumn,secnumarabic,superscriptaddress,amssymb, nobibnotes, ap
s, prl, reprint]{revtex4-1}
\usepackage{amsmath}
\usepackage{graphicx}

\begin{document}

\title{Diffractive Electron-Nucleus Scattering
  and Ancestry in Branching Random Walks}%

\author{A. H. Mueller}
\affiliation{Department of Physics, Columbia University, New York,
  New York 10027, USA}
\author{S. Munier}
\affiliation{Centre de physique th\'eorique, \'Ecole polytechnique, CNRS,
  Universit\'e Paris-Saclay, 1 route de Saclay, 91128 Palaiseau, France}
\date{August 12, 2018}%

\begin{abstract}
  We point out an analogy between diffractive
  electron-nucleus scattering events
  and realizations
  of one-dimensional branching random walks selected according to
  the height of the genealogical tree
  of the particles near their boundaries.
  This correspondence is made transparent in an event-by-event
  picture of diffraction, emphasizing the statistical
  properties of gluon evolution,
  from which new quantitative predictions straightforwardly
  follow: we are able to determine the
  distribution of the total invariant mass produced diffractively,
  which is an interesting observable
  that can potentially be measured at a future electron-ion
  collider.
\end{abstract}
\maketitle

\emph{Introduction.} Diffraction is an elementary consequence of the particle-wave duality
postulated by quantum mechanics. Therefore, diffractive patterns are expected
to be observed in the scattering of elementary particles off extended
objects such as hadrons or nuclei.
However, the microscopic interpretation of diffraction
turns out to be subtle. Indeed, it is well known that
nuclei are loose compounds
 of hadrons, which themselves appear as fragile bound states
of quarks as soon as they are involved in collisions at
center-of-mass energies much larger than typically
the mass energy of a nucleon.
Naively, an energetic electron
colliding with a hadron or nucleus,
a process known as ``deep-inelastic scattering'' (DIS),
would knock out a quark in each scattering event;
then, as a consequence of confinement, the final state
would almost systematically consist of many new hadrons
distributed all over the detector.

But this is not at all what has been seen experimentally.
Indeed, one of the outstanding results of the
DESY-HERA electron-proton
collider is the observation of a significant fraction of the
events (about 10\%) in which the scattered proton is left intact
and is surrounded by an angular region of variable size, empty of
particles that we shall call ``gap."
What has been observed in electron-proton collisions
should also happen in electron-nucleus scattering.
Testing whether this expectation is true can be
achieved at a future electron-ion collider.

Diffraction in DIS on protons has been studied extensively,
both experimentally~(for a review, see~\cite{Schoeffel:2009aa})
and theoretically (see~\cite{Kovchegov:2012mbw}
and references therein).
But its quantitative theoretical description
in the framework of the established theory of the strong interaction, quantum
chromodynamics (QCD),
remains a challenge. While
it is known that
the total diffractive cross section can be explained
economically and elegantly by saturation models~\cite{GolecBiernat:1998js},
little analytical insight has been gained for more exclusive
diffractive observables.

In this Letter, we focus on the diffractive events in deep-inelastic
scattering off a large nucleus
in which the nucleus is left intact,
but a hadronic state of large invariant mass $M_X$ is nevertheless
produced. We explain how to characterize them microscopically, and we show
that these hadrons are generated from a similar mechanism as
the common ancestor of a set of particles at the frontier of
a one-dimensional branching random walk.
We deduce from this very mechanism a simple analytical prediction,
Eq.~(\ref{eq:distrib_y0}) below,
which we test
against the numerical integration of a previously known
equation governing the energy dependence of high-mass diffraction.

\begin{figure}[t]
\includegraphics[width=.4\textwidth]{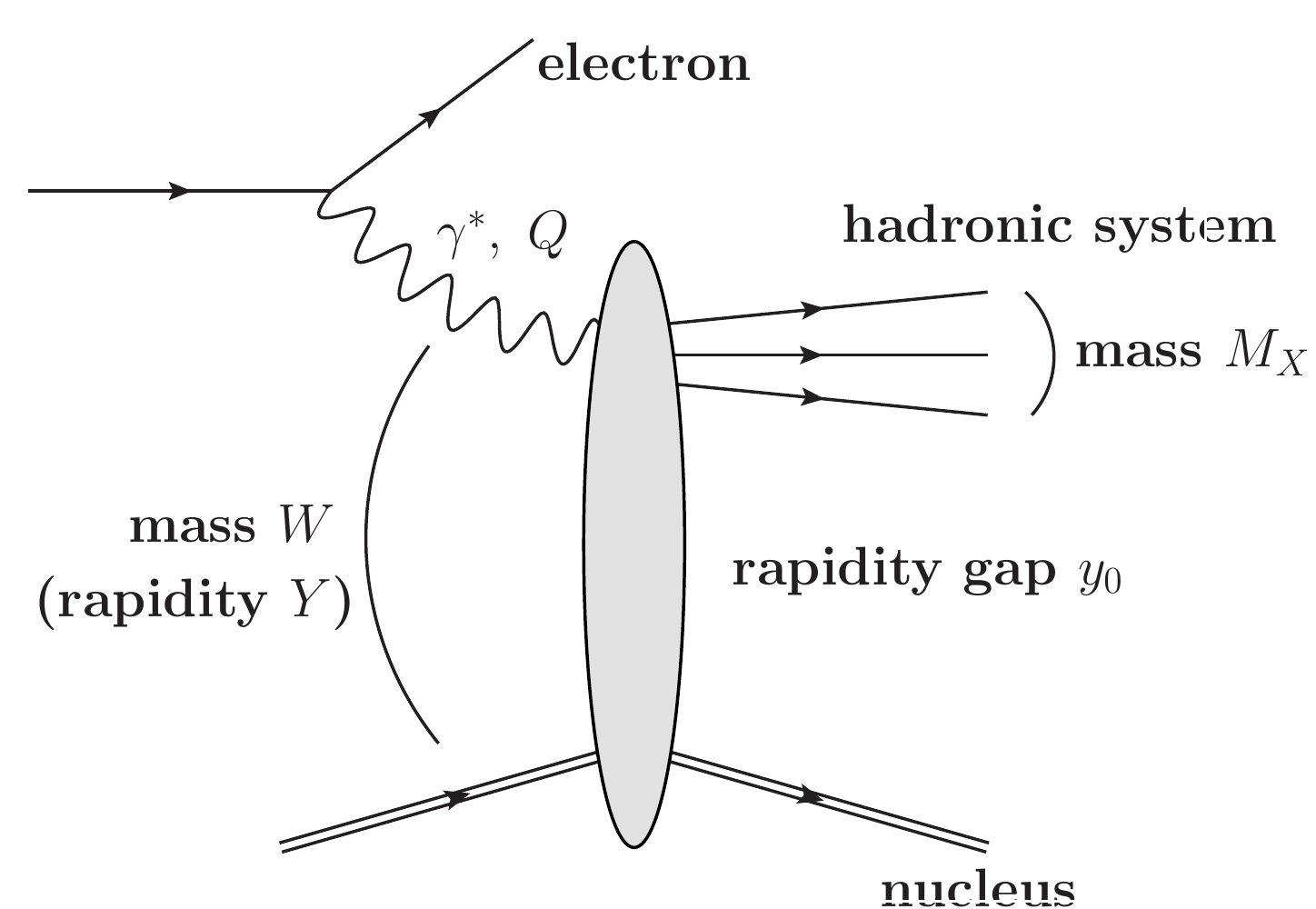}
  \caption{\label{fig:ddis}
    Schematic
    representation
    of a diffractive event. The initial-state particles are incoming from
    the left, the final state is on the right.
    The interaction of the electron with the
    nucleus is mediated by a virtual photon.
    While the nucleus is transferred unaltered in its nature
    to the final state, the photon converts to a set of hadrons of
    total invariant mass $M_X$. The rapidity gap is an angular region
    around the nucleus in which no particle is observed.
  }
\end{figure}

\emph{Picture of electron-nucleus scattering at
  high energy.} The scattering of the electron off the nucleus
necessarily proceeds through the exchange
of a virtual photon $\gamma^*$. We shall denote its virtuality by $Q$
and  the center-of-mass energy
of the $\gamma^*$--nucleus subprocess by $W$.
These variables are enough to label the total cross section.
In the case of diffractive scattering (see Fig.~\ref{fig:ddis}),
the cross section also
depends on the total invariant mass $M_X$ of the produced hadrons.
It is convenient to use, instead of $M_X$,
the dimensionless variable $\beta=Q^2/(Q^2+M_X^2)$, in terms of which
the DIS experiments traditionally present
diffractive data~\cite{Schoeffel:2009aa},
or the logarithm of its inverse $\tilde y_0\equiv \ln 1/\beta$.
The gap can then be characterized
by the Lorentz-invariant rapidity variable
$y_0\equiv Y-\tilde y_0$, where
$Y\equiv\ln 1/x_{Bj}$,
with $x_{Bj}=Q^2/(Q^2+W^2)$, is the total relative rapidity
of the photon with respect to the nucleus.

When the energy of the reaction is large, it is possible
to choose a reference frame
in which the photon is fast enough to almost always convert
to a quark-antiquark pair (which we shall call ``onium'')
before interacting. For our purpose, the only
relevant parameter to characterize the latter is the
distance $r$ between the trajectories of the quarks, which
 can be considered unchanged throughout a scattering
 at high relative rapidity. The
distribution of $r$ for a given photon virtuality follows
from simple electrodynamics.
Hence electron-nucleus scattering
is tantamount to onium-nucleus scattering.
A scattering event
occurs as soon as at least one gluon is exchanged
between the onium and the nucleus.

A nucleus is \textit{a priori} a very complicated composite object.
However, a \textit{large} nucleus is made of many hadrons, which
can be considered uncorrelated. Considering, furthermore, the number~$N_c$
of colors to be a large parameter, the rapidity evolution of the forward
elastic amplitude $T(r,y)$ for the
scattering of the onium off the nucleus can be established within QCD
in these limits.
It is given by the Balitsky-Kovchegov (BK)
equation~\cite{Balitsky:1995ub}
\begin{equation}
  \frac{\partial T(r,y)}{\partial y}
  =\bar\alpha \left[
    \chi T(r,y)
    -T{\otimes}T(r,y)
    \right],
  \label{eq:BK}
\end{equation}
where $\bar\alpha$ is proportional to the
product of the strong coupling constant $\alpha_s$
by the number of colors, $\bar\alpha=\alpha_s N_c/\pi$;
$\chi$ in the first term
is the linear operator
that acts on a function~$f$ of~$r$ as
\begin{equation*}
\chi f(r)=\int\frac{d^2 r'}{2\pi}
  \frac{r^2}{r^{\prime 2}(r-r^\prime)^2}
  [f(r')+f(r-r')-f(r)];
\end{equation*}
and finally, the second term in the rhs of~(\ref{eq:BK}) is the convolution
\begin{equation*}
f{\otimes}f(r)=\int\frac{d^2 r'}{2\pi}
  \frac{r^2}{r^{\prime 2}(r-r^\prime)^2}
f(r')f(r-r').
\end{equation*}
The elastic onium-nucleus scattering cross section per unit surface~\cite{Note1}
is $\sigma_\text{el}=T^2$ (since $T$ is essentially real at high energy)
evaluated at rapidity $y=Y$,
and the total cross section is
twice $T$ as a consequence of the optical theorem:
$\sigma_\text{tot}=2T$. (The total electron-nucleus cross section may
then easily be calculated from $\sigma_\text{tot}$.)
Thanks to these notations, the structure of the BK equation~(\ref{eq:BK})
is quite clear. The first term, linear in $T$, encodes the rise of the
amplitude due to the multiplication of the gluons
in the state of the onium as the rapidity is increased, i.e., as
shorter-lived quantum fluctuations become relevant for the scattering.
It is well known that in the large-$N_c$ limit and in a light cone gauge,
the Fock state of an onium can conveniently
be represented by a set of dipoles
of various sizes, and rapidity evolution
can be thought of as a cascade of independent
$1\rightarrow 2$ splittings of color dipoles~\cite{Mueller:1993rr}.
A light cone perturbation theory calculation in the framework
of QCD leads to the expression of
the splitting probability density of a dipole of size $r$
into dipoles of sizes $r'$, $r-r'$ as its rapidity is increased by $dy$;
it reads
\begin{equation*}
dp(r\rightarrow r',r-r')=
\bar\alpha dy\frac{d^2 r'}{2\pi}\frac{r^2}{{r'}^2 (r-r')^2}.
\end{equation*}
The operator $\chi$, which is constructed from the integral of
$({1}/{\bar\alpha}){dp}/{dy}$,
is also the kernel of the evolution equation solved by the
mean number density $n(r,y)$ of dipoles at rapidity $y$ in
an onium of initial size $r$: $\partial_y n=\bar\alpha\chi n$,
which is nothing but
the Balitsky-Fadin-Kuraev-Lipatov
equation~\cite{Lipatov:1976zz}.
The second term in the BK equation~(\ref{eq:BK}),
significant only when $T={\cal O}(1)$,
keeps the amplitude unitary ($T\leq 1$) throughout the evolution.

It is useful to note that the BK equation is in essence similar to
the {nonlinear diffusion equation}, known
as the Fisher-Kolmogorov-Petrovsky-Piscounov (FKPP)
equation~\cite{doi:10.1111/j.1469-1809.1937.tb02153.x}:
these two equations actually belong to the same universality class~\cite{Note2}.
Starting from this correspondence, one
can take advantage of the available mathematical knowledge on
the FKPP equation (for a review, see Ref.~\cite{VANSAARLOOS200329}).
One knows that, for a vast class of initial conditions,
its solution converges
to a traveling wave at large~$y$, namely, a front
connecting $T=1$ for $r$ large to $T=0$ for $r$ small, the rapidity evolution
of which consists of a mere translation in $r$.
The transition region is located around a rapidity-dependent size
$r_s(y)$ related to the saturation momentum $Q_s$ by
$r_s=1/Q_s$. The analytical expression of $Q_s(y)$ for $\bar\alpha y\gg 1$ reads
\begin{equation}
  Q_s^2(y)=Q_\text{MV}^2\frac{e^{\bar\alpha y \chi'(\gamma_0)}}
  {(\bar\alpha y)^{3/2\gamma_0}},
  \label{eq:Qs}
\end{equation}
up to a multiplicative constant of order one
depending on the very definition of $Q_s$.
The complex function $\chi(\gamma)$ is the set of the eigenvalues
of the $\chi$ operator associated with its eigenfunctions $r^{2\gamma}$,
and $\gamma_0$ solves $\chi(\gamma_0)=\gamma_0\chi'(\gamma_0)$.
Explicitly, $\chi(\gamma)=2\psi(1)-\psi(\gamma)-\psi(1-\gamma)$,
where $\psi$ is the digamma function,
and $\gamma_0\simeq 0.63$.
Equation~(\ref{eq:Qs}) holds whenever
the initial condition falls fast enough as~$r$ decreases. More
precisely, if $T(r,0)\underset{r\rightarrow 0}{\sim} r^{2\lambda}$,
then $\lambda$ must be larger than $\gamma_0$~\cite{MR705746}.

An analytical expression for the asymptotic shape of the front is
also known. It reads
\begin{equation}
  T(r,y)=c_T \ln\frac{1}{r^2Q_s^2(y)}\left[r^2Q_s^2(y)\right]^{\gamma_0},
  \label{eq:T}
\end{equation}
where $c_T$ is a numerical constant.
This equation is valid when $T\ll 1$,
and in the so-called
scaling region~\cite{Stasto:2000er}.
These two conditions translate
into the inequalities
$1\ll|\ln r^2Q_s^2(y)|\ll\sqrt{\chi''(\gamma_0)\bar\alpha y}$.
Throughout, we will always assume that
$r$ is such that both these inequalities are fulfilled.

The initial condition for $T$
describes the interaction amplitude of the onium with the
nucleus at low energy.
A nucleus in its rest frame is a dense system of quarks.
In the so-called McLerran-Venugopalan (MV) model~\cite{McLerran:1993ni},
it is characterized by a momentum scale $Q_\text{MV}$ function
of the atomic number.
(Its value is of order 1~GeV for heavy nuclei such as lead or gold).
The scattering amplitude of an onium with a nucleus
may be approximated by $T(r,y=0)=1-e^{-{r^2Q_\text{MV}^2}/{4}}$.
Onia of size much larger than typically $r_\text{MV}\equiv 1/Q_\text{MV}$
are absorbed, while the nucleus appears transparent
to onia of size much smaller than $r_\text{MV}$.
We note that $T(r,y=0)\sim r^2$ for small $r$:
hence, the solutions~(\ref{eq:Qs}) and (\ref{eq:T}) indeed apply.

The BK equation~(\ref{eq:BK})
is also an equation for the $\tilde y$ evolution of the
probability $P(r,\tilde y|R)$ that there be at least a dipole larger
than some $R$ in an onium of initial size $r$~\cite{Mueller:2014fba}.
The initial condition in this case
reads $P(r,\tilde y=0|R)=\theta(\ln r^2/R^2)$,
which is of course ``steep enough'' for the
asymptotic solution~(\ref{eq:T}) to be valid. Then, thanks to
the universality properties of the asymptotic solution to the BK equation,
$P(r,\tilde y|R)$ has the same expression as~$T(r,y)$ in Eq.~(\ref{eq:T})
up to the substitutions $y\leftrightarrow \tilde y$ and
$Q_\text{MV}\leftrightarrow 1/R$ (that appears in Eq.~(\ref{eq:Qs}))
and maybe up to the overall normalization constant.
Thus, one can write $\sigma_{\text{tot}}\propto P(r,Y|1/Q_\text{MV})$.

\emph{Diffraction from rare fluctuations.} If $r$ is small compared to $1/Q_s(Y)$---i.e., the onium
is far from the saturation region of the nucleus---then, from Eq.~(\ref{eq:T}),
$T$ is small.
Since the forward elastic amplitude~$T$
is related to the total, elastic, and inelastic cross sections through
\begin{equation*}
  \sigma_\text{tot}=2T,\quad
  \sigma_\text{el}=T^2,\quad
  \sigma_\text{in}=\sigma_\text{tot}-\sigma_\text{el},
\end{equation*}
one sees that $\sigma_\text{el}$ is of second order in $T$, while
$\sigma_\text{in}$ is of first order, and thus dominates
$\sigma_\text{tot}$.

\begin{figure}[t]
  \begin{tabular}{c|c}
  \includegraphics[width=.198\textwidth]{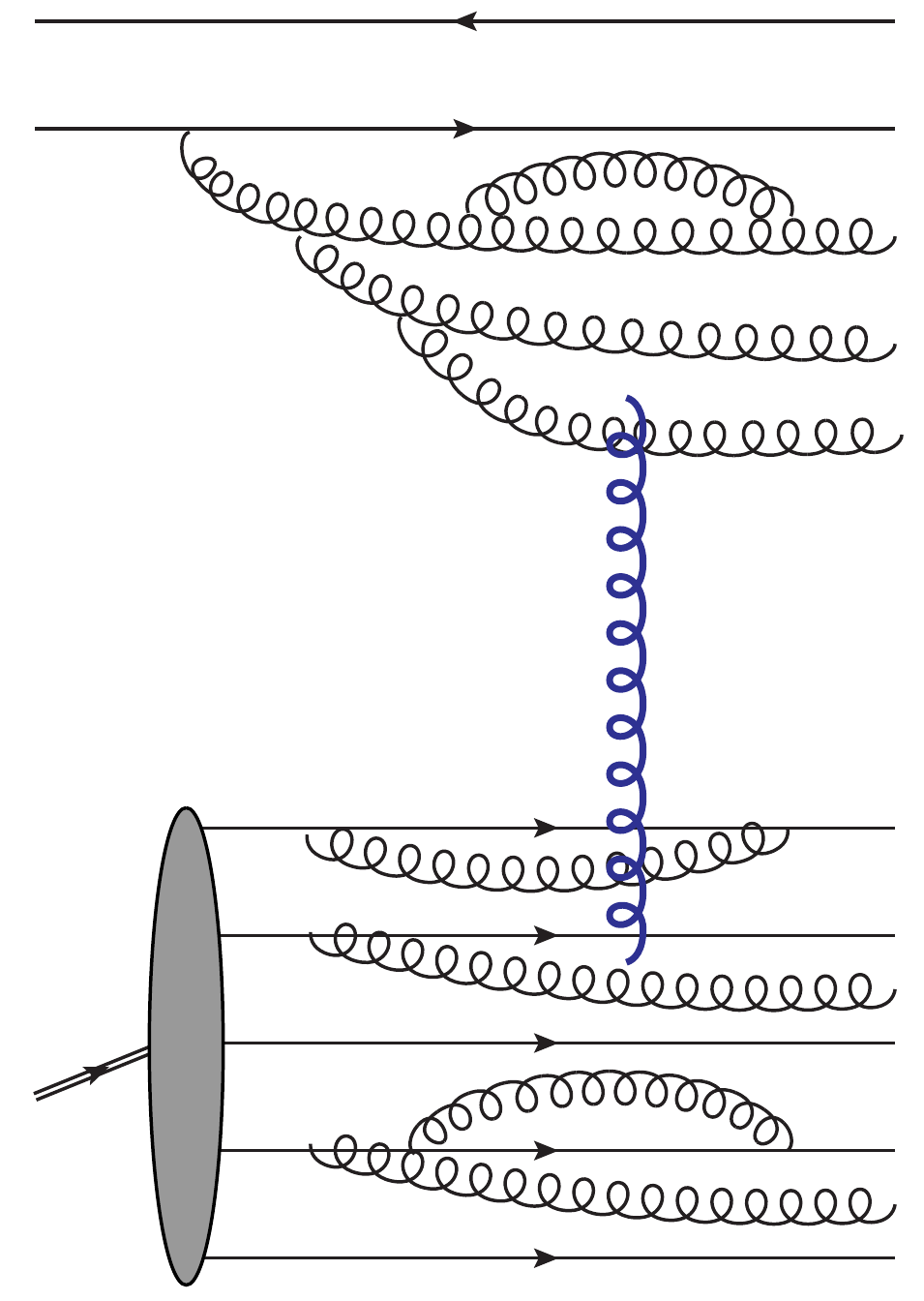}\ \ &\ \
  \includegraphics[width=.22\textwidth]{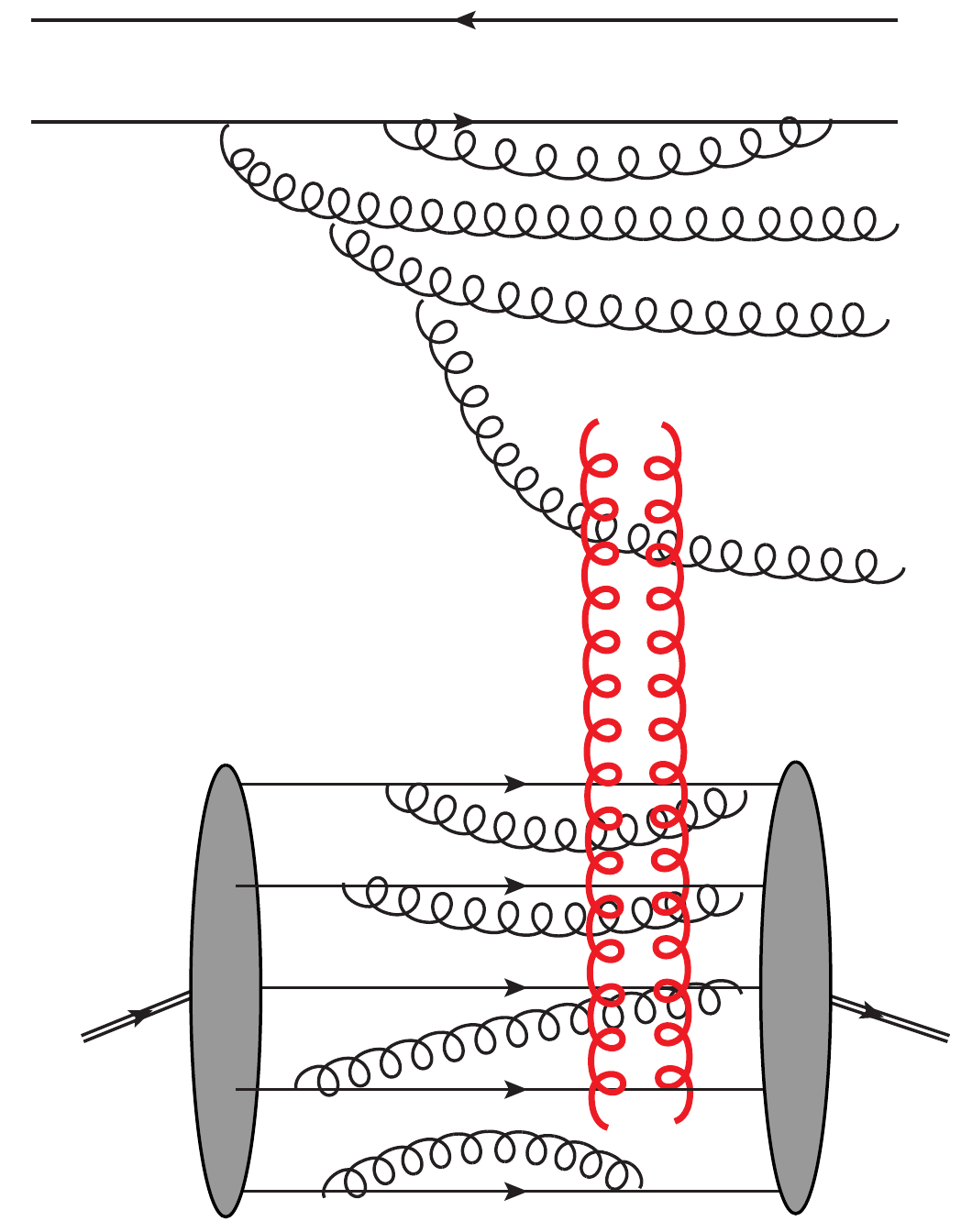}\\
  (a) & (b)
  \end{tabular}
  \caption{\label{fig:y0-frame}
    Light cone perturbation theory diagrams contributing to onium-nucleus
    scattering in a frame in which
    the nucleus carries the rapidity $y_0$ and
    the onium $\tilde y_0=Y-y_0$.
    (In the large-$N_c$ limit, the gluons are replaced by 0-size $q\bar q$ pairs,
    and the onium Fock state consists of a set of dipoles.)
    (a) Nondiffractive scattering.
    (b) Diffractive event
    with a rapidity gap $y_0$.
    In case (a), the scattering
    consists of the exchange of [most probably, if $r\ll 1/Q_s(Y)$] one
    gluon, which results in a breakup of
    the nucleons and thus of the nucleus. In case (b), one or multiple
    pairs of gluons grouped
    in color singlet states are exchanged, in which case the nucleus scatters
    elastically and no gluon is emitted
    in the direction of the nucleus momentum.
}
\end{figure}

A diffractive event can occur
with non-negligible probability only if
a large dipole occurs in the Fock state of the onium at
some point in the evolution.
Indeed, only for such realizations of the evolution
the scattering amplitude can be of order~1, and elastic
scattering processes are thus probable.
(Examples of Feynman diagrams contributing to the onium-nucleus
diffractive vs total amplitudes
are shown in Fig.~\ref{fig:y0-frame}.)
Assume that such a dipole of size $r_0$ is produced at rapidity
$\tilde y_0=Y-y_0$. The condition that the whole partonic system
scatters elastically
with a significant probability is that $r_0$ be larger than the
inverse saturation scale of the nucleus {evaluated at the
  rapidity $Y-\tilde y_0=y_0$}: $r_0>1/Q_s(y_0)$.
Such an event will exhibit a rapidity gap of size $y_0$.

From this picture, we can immediately infer that
the diffractive cross section conditioned to a given rapidity gap~$y_0$
is tantamount to the probability $P\boldsymbol(r,\tilde y_0|1/Q_s(y_0)\boldsymbol)$.
As discussed above, the latter is given by the solution to
the BK equation~(\ref{eq:T}) up to the appropriate substitution of the
variables and parameters
\begin{equation*}
  \frac{d\sigma_\text{diff}}{dy_0}
  =c_\text{diff}\ln\frac{1}{r^2\tilde Q_s^2(\tilde y_0)}
  \left[r^2\tilde Q_s^2(\tilde y_0)\right]^{\gamma_0},
\end{equation*}
where $c_\text{diff}$ is a constant, and
the momentum $\tilde Q_s$ reads
\begin{equation*}
  \tilde Q_s^2(\tilde y_0)=Q_s^2(y_0)\frac{e^{\bar\alpha \tilde y_0\chi'(\gamma_0)}}
  {(\bar\alpha \tilde y_0)^{3/2\gamma_0}}.
\end{equation*}

A straightforward calculation [using Eqs.~(\ref{eq:Qs}) and (\ref{eq:T})]
leads to our main result
for the distribution of the size of the rapidity gap.
The simplest expression is obtained for the differential
diffractive cross section normalized by the total cross section
\begin{equation}
\frac{1}{\sigma_{\text{tot}}}\frac{d\sigma_\text{diff}}{dy_0}
\propto\left(\frac{\bar\alpha Y}{\bar\alpha y_0\,\bar\alpha(Y-y_0)}
  \right)^{3/2}
\label{eq:distrib_y0}
\end{equation}
in such a way that the overall coefficient, which
we have not been able to determine, is a pure number independent of
the parameters $r$ and $Y$.
The formula~(\ref{eq:distrib_y0})
is valid whenever $\bar\alpha y_0$ is distant from its two boundaries
at zero and $\bar\alpha Y$ by more than typically one unit.
We also recall that this result is an asymptotic
limit for $\bar\alpha Y$ large and that it holds when the size
$r$ of the onium is picked in the
scaling region.

\emph{Genealogy in branching random walks.} Our whole discussion of the structure of diffractive events
turns out to be parallel to the discussion of the genealogy of
particles near the boundary of a
branching random walk (BRW) in Ref.~\cite{0295-5075-115-4-40005}.

Consider a BRW in time~$t$ and in
the real variable~$x$,
starting with one single particle,
defined with the help of a stochastic process such
that the mean density of particles $n(x,t)$ obeys the equation
$\partial_t n=\chi n$; $\chi$ is an appropriate operator acting
on $n$ viewed as a function of $x$ and encoding the
microscopic process: for example, $\chi=\partial_x^2+1$.
$\chi$ admits the eigenfunctions $e^{-\gamma x}$ and we denote by
$\chi(\gamma)$ the corresponding eigenvalues.
After the (large) time~$t$, pick exactly two particles,
choosing them either
{(i)} according to the Boltzmann weight $e^{-\lambda x}$
(i.e., the particle number $j$ sitting at position $x_j$ at time $t$
is picked with probability
$e^{-\lambda x_j}/\sum_k e^{-\lambda x_k}$)
or {(ii)} to be exactly the two leftmost particles,
and look for the first common ancestor splitting time $t-t_0$.
Then, according to Ref.~\cite{0295-5075-115-4-40005},
$t_0$ is distributed as
\begin{equation}
p(t_0)=
c_p\left(\frac{t}{t_0(t-t_0)}\right)^{3/2},
\ \text{with}\ c_p=\frac{1}{\bar\gamma}\frac{1}{\sqrt{2\pi\chi''(\gamma_0)}},
\label{eq:p}
\end{equation}
where $\bar\gamma=\lambda$ in case {(i)} if $\lambda>\gamma_0$, and
$\bar\gamma=\gamma_0$ in case {(ii)}.
$\gamma_0$ solves $\chi(\gamma_0)=\gamma_0\chi'(\gamma_0)$.

In the same way as in our diffraction calculation,
the common ancestor of the boundary
particles also corresponds to a fluctuation,
in the form of a particle sent to the left of the \textit{expected}
position of the leftmost particle,
occurring
in the course of the evolution at time $t-t_0$.
Hence, the two problems are intimately related: up to the overall
normalization, which is determined in the case of the genealogies, but
not in the case of diffraction,
$({1}/{\sigma_\text{tot}})({d\sigma_\text{diff}}/{dy_0})$
corresponds to $p(t_0)$,
with the identifications $\bar\alpha Y\leftrightarrow t$,
$\bar\alpha y_0\leftrightarrow t_0$.

\emph{Numerical test.} An equation for the diffractive cross section
with a rapidity gap $y_0$
was established some time ago in QCD by Kovchegov
and Levin (KL)~\cite{Kovchegov:1999ji}
(see also Refs.~\cite{Kovchegov:2012mbw,Kovner:2001vi}).
It can be put in the form of two appropriately matched
evolution equations in the total rapidity variable $y$, which
both turn out to be of the BK type.
While this formulation has not led to much analytical insight,
in particular, for the
gap distribution we are addressing here, it is very convenient
for the numerical computation of
the diffractive cross section, since the BK equation is easily
discretized, implemented, and solved using standard
algorithms~\cite{Levin:2001yv}.

\begin{figure}
\includegraphics[width=.5\textwidth]{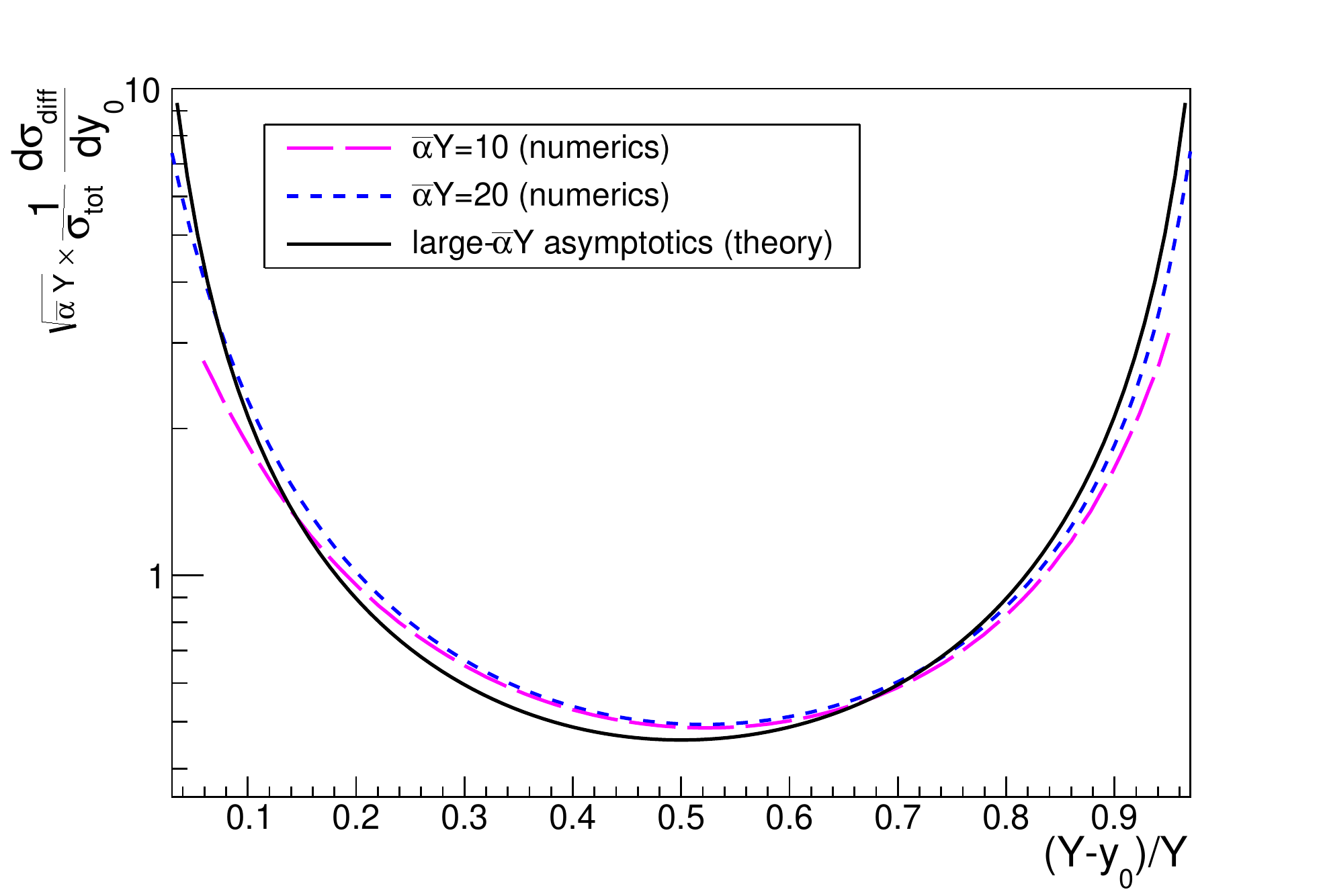}
  \caption{\label{fig:numerics}Rescaled distribution of $(Y-y_0)/Y$
    calculated from the numerical
    integration of the KL equation
    for two different values of the total rapidity, compared
    to the asymptotic theoretical formula~(\ref{eq:distrib_y0}), with
    an \textit{ad hoc} global normalization factor.
  The onium size is chosen to be well in the scaling region. (See the
  main text for the details).
}
\end{figure}

We have
computed the rapidity-gap distribution for two values of the total rapidity,
$\bar\alpha Y=10$ and $20$.
(These rapidities are of course too large to be realistic for phenomenology,
but our goal here is to test our asymptotic prediction.)
We have chosen $r$ in such a way that $|\ln r^2Q_s^2(Y)|\simeq 7.2$,
comfortably in the scaling region in both cases.
The results are  presented in Fig.~\ref{fig:numerics} and compared
to the analytical prediction~(\ref{eq:distrib_y0})
in appropriately rescaled variables chosen such that
the expected asymptotic distribution be independent of $\bar\alpha Y$.
The overall coefficient of the latter is not predicted in our approach.
We could fit it to the numerical data, but
interestingly enough, just setting it to be
that predicted for BRW, namely, $c_p$
in Eq.~(\ref{eq:p}), with $\bar\gamma=1$, leads to a
remarkably good agreement between the numerical data
and the prediction.

\emph{Conclusion.} We have found that the distribution of the size $y_0$
of rapidity gaps in diffractive onium-nucleus scattering
can be calculated analytically for fixed
large-center-of-mass energies.
Surprisingly enough, our quantitative prediction follows
quite straightforwardly from simple
considerations on the mechanism
how the Fock state of a quark-antiquark pair evolves
when one 
increases its rapidity. The essence of this evolution is that of
a one-dimensional
branching random walk, and viewed in such a picture,
diffractive events are due to the
existence of a large fluctuation in the evolution.
The rapidity at which it occurs
determines the size of the gap.

This large fluctuation can also be identified with the common ancestor
of a few extreme objects generated by the BRW.
The latter problem is of interest in the study of disordered systems.
It was known before that the BK and FKPP equations are in the
same universality class~\cite{Munier:2003vc},
and also that the energy evolution of the scattering amplitude of
ultrahigh-energy hadrons
may be analogous to the time evolution of a reaction-diffusion
process, the evolution of which is
described by an equation belonging to
the universality class of the {stochastic} FKPP
equation~\cite{Iancu:2004es}.
But to our knowledge, this is the first time that the
statistical properties of genealogical trees prove
of direct relevance in the context of particle or nuclear physics.
Hence, our Letter contributes to
bridge \textit{a priori} unrelated fields of physics.

The results we have obtained here can be converted into
new predictions for the mass distribution in diffractive virtual photon-nucleus
scattering $\beta d\sigma^{\gamma^* A}_\text{diff}/d\beta$,
measurable at a future
electron-ion collider. At fixed $W$ and $Q$, the latter is actually identical to
the distribution of rapidity gaps
$d\sigma^{\gamma^* A}_\text{diff}/dy_0$, which can be calculated
by convoluting the onium cross section
$d\sigma_\text{diff}/dy_0$ with the known distribution
of the sizes~$r$ of quark-antiquark pairs in the Fock state of the virtual photon
(see, e.g.,~\cite{GolecBiernat:1998js}).
This is actually straightforward when the photon is polarized longitudinally,
since in this case, the distribution
is peaked around the inverse photon virtuality; i.e., $r$
can essentially be identified to $1/Q$. For transversely polarized photons,
since the $r$ distribution corresponding to a given $Q$ is wider, a
better knowledge of $d\sigma_\text{diff}/dy_0$ outside of the scaling region
would be needed. Further developments, along with more
numerical studies, can be found in Ref.~\cite{Mueller:2018ned}.

\acknowledgements{
The  work  of  A. H. M.  is  supported in part by
the U.S. Department of Energy Award No. DE-FG02-92ER40699.
The work of S. M. is supported in part by the Agence Nationale
de la Recherche under Project No. ANR-16-CE31-0019.
We thank Bernard Pire for urging us to try to make the material
of the present Letter accessible to a wider audience
and for useful suggestions on the manuscript.
}


\begin{thebibliography}{35}%
\bibitem{Schoeffel:2009aa}%
  L.~Schoeffel, %
{Prog. Part. Nucl. Phys. \textbf{65}, 9 (2010)}.%
\bibitem{Kovchegov:2012mbw}%
  Y.~V. Kovchegov and E.~Levin,
  {\emph{{Quantum Chromodynamics at High Energy}}} ({Cambridge University
  Press}, Cambridge, England, 2012),   Vol.~33.%
\bibitem{GolecBiernat:1998js}%
  K.~J. Golec-Biernat and M.~Wusthoff, %
{Phys. Rev. D \textbf{59}, 014017 (1998)}.%
\bibitem{Balitsky:1995ub}%
  I.~Balitsky, %
{Nucl. Phys. \textbf{B463}, 99 (1996)};
  Y.~V. Kovchegov, %
{Phys. Rev. D \textbf{61}, 074018 (2000)}.%
\bibitem{Note1}%
  Throughout, we focus on the diffractive cross section
  {per unit surface} in the transverse plane, thus evaluated at a
  fixed impact parameter $b$. To obtain the cross section integrated over $b$,
  one would need a model for the $b$ dependence of $Q_\protect \text{MV}$ (see,
  e.g.,~\cite{Kowalski:2003hm} in the case of the
  proton). One may also simply assume $Q_\protect \text{MV}$ constant over the
  surface of the nucleus, as was done for DIS off the proton in Ref.~\cite{GolecBiernat:1998js}, in which case the integration over~$b$ just amounts to
  multiplying by the cross section surface of the nucleus.%
\bibitem{Mueller:1993rr}%
  A.~H. Mueller, %
{Nucl. Phys. \textbf{B415}, 373 (1994)}.%
\bibitem{Lipatov:1976zz}%
  L.~N. Lipatov,  {Sov.
  J. Nucl. Phys. \textbf{23}, 338
  (1976)};
  E.~A. Kuraev, L.~N. Lipatov,
   and V.~S. Fadin,    {Sov. Phys. JETP   \textbf{45}, 199 (1977)};
  I.~I. Balitsky and L.~N. Lipatov,  {Sov.
  J. Nucl. Phys. \textbf{28}, 822
  (1978)}.%
\bibitem{doi:10.1111/j.1469-1809.1937.tb02153.x}%
  R.~A. Fisher, 
{Ann. Eugenics \textbf{7}, 355 (1937)};
  A.~Kolmogorov, I.~Petrovsky,  and N.~Piscounov,  {Bull. Univ. \'Etat Moscou \textbf{A1}, 1 (1937)}.%
\bibitem{Munier:2003vc}%
  S.~Munier and R.~B. Peschanski, %
{Phys. Rev. Lett. \textbf{91}, 232001 (2003)}.%
\bibitem{Note2}%
  An exact mapping can be exhibited for the Fourier transform
  of $T$ defined as $\tilde T(k,y)=\int d^2r/(2\pi r^2)e^{ikr}T(r,y)$
  and in the so-called diffusive approximation; see Ref.~\cite{Munier:2003vc}
\bibitem{VANSAARLOOS200329}%
  W.~van
  Saarloos,
  {Phys. Rep. \textbf
  {386}, 29  (2003)}.%
\bibitem{MR705746}%
  M.~Bramson,  {Mem.
  Amer. Math. Soc. \textbf{44}, iv+190 (1983)}.%
\bibitem{Stasto:2000er}%
  A.~M. Stasto, K.~J. Golec-Biernat,  and J.~Kwiecinski, %
{Phys. Rev. Lett. \textbf{86}, 596 (2001)};
  E.~Iancu, K.~Itakura,    and L.~McLerran,  {Nucl. Phys. \textbf{A708}, 327 (2002)};
  A.~H. Mueller and D.~N. Triantafyllopoulos,
  {Nucl. Phys.  \textbf{B640}, 331 (2002)}.%
\bibitem{McLerran:1993ni}%
  L.~D. McLerran and R.~Venugopalan, %
{Phys. Rev. D \textbf{49}, 2233 (1994)}.%
\bibitem{Mueller:2014fba}%
  A.~H. Mueller and S.~Munier, %
{Phys. Lett. B \textbf{737}, 303 (2014)}.%
\bibitem{0295-5075-115-4-40005}%
  B.~Derrida and P.~Mottishaw,
  {Europhys. Lett.   \textbf{115}, 40005 (2016)}.%
\bibitem{Kovchegov:1999ji}%
  Y.~V. Kovchegov and E.~Levin, %
{Nucl. Phys. \textbf{B577}, 221 (2000)}.%
\bibitem{Kovner:2001vi}%
  A.~Kovner and U.~A. Wiedemann, %
{Phys. Rev. D \textbf{64}, 114002 (2001)};
  M.~Hentschinski, H.~Weigert,  and A.~Schafer, %
{Phys. Rev. D \textbf{73}, 051501 (2006)};
  Y.~Hatta, E.~Iancu,
  C.~Marquet, G.~Soyez,  and D.~N. Triantafyllopoulos, 
{Nucl. Phys. \textbf{A773}, 95
  (2006)}.%
\bibitem{Levin:2001yv}%
  E.~Levin and M.~Lublinsky, %
{Phys. Lett. B \textbf{521}, 233 (2001)};
{Eur. Phys. J. C \textbf{22},   647 (2002)};
 {Nucl. Phys. \textbf{A712}, 95 (2002)}.%
\bibitem{Iancu:2004es}%
  E.~Iancu, A.~H. Mueller,
   and S.~Munier,  {Phys. Lett. B \textbf{606}, 342 (2005)};
  E.~Iancu and D.~N. Triantafyllopoulos,
  {Nucl. Phys. \textbf{A756}, 419 (2005)}.%
\bibitem{Mueller:2018ned}%
  A.~H. Mueller and S.~Munier, Phys. Rev. D (2018) (to appear),
arXiv:1805.02847.
\bibitem{Kowalski:2003hm}%
  H.~Kowalski and D.~Teaney, %
{Phys. Rev. D \textbf{68}, 114005 (2003)};
H.~Kowalski, L.~Motyka,    and G.~Watt,
{Phys. Rev. D \textbf{74}, 074016 (2006)};
  G.~Watt and H.~Kowalski, %
{Phys. Rev. D \textbf{78}, 014016 (2008)}.%
\end{thebibliography}

%

\end{document}